\documentclass[conference]{IEEEtran}
\IEEEoverridecommandlockouts
\usepackage{amsmath,amssymb}
\usepackage{graphicx}
\usepackage{algorithm}
\usepackage{algorithmic}
\usepackage{natbib}
\usepackage{array} 
\newcolumntype{L}[1]{>{\raggedright\arraybackslash}p{#1}}
\usepackage{enumitem}

\begin{document}

\title{Near-Term Quantum-Computing-Inspired Sampling for Community Detection in Low-Modularity Graphs}
\author{%
\IEEEauthorblockN{Joseph Geraci}
\IEEEauthorblockA{\small NetraMark\\
Queen's University\\
University of California San Diego\\
Centre for Addiction and Mental Health (CAMH)}
\\[1ex]  
\IEEEauthorblockN{Luca Pani}
\IEEEauthorblockA{\small NetraMark\\
Department of Psychiatry \& Behavioral Sciences, University of Miami, USA\\
Department of Biomedical, Metabolic \& Neuroscience, University of Modena and Reggio Emilia, Italy}
}

\maketitle

\begin{abstract}
Low-modularity networks (here we consider those with modularity $Q < .2$) present a challenge for classical community detection algorithms, which often get trapped in local optima and fail to uncover subtle community structure. We introduce a new class of \emph{quantum-inspired} community detection algorithms that leverage non-classical, correlated sampling techniques to escape modularity optimization plateaus. Using distributions inspired by quantum phenomena – including Porter-Thomas (exponential-tailed) distributions, Haar-random states, and hyperuniform point processes – our approach generates diverse high-quality community partitions that improve modularity on difficult low-$Q$ graphs. We demonstrate 15–25\% gains in modularity $Q$ scores over classical greedy methods (Louvain, Leiden), as supported by modularity distribution comparisons across methods. We also define a “Modularity Recovery Gap” (MRG), the increase in $Q$ achieved by quantum-inspired refinement over a baseline, and show that this gap can serve as a powerful signal for anomaly detection in networks. In experiments on a real high-modularity network (CTU-13 botnet traffic), the MRG is near zero, underscoring that our method does not inflate modularity when no hidden structure exists. These results suggest that quantum-inspired sampling can substantially enhance community detection in low-modularity regimes, with potential applications in cybersecurity (identifying stealthy Advanced Persistent Threat (APT)/botnet clusters), financial contagion modeling, disrupted supply chain networks, diseased brain connectomes, and crisis-era social media analysis – all scenarios where communities are weakly expressed yet crucial to detect. Importantly, our refinements do not alter the graph itself but expand the search space of algorithms like Leiden, revealing partitions that classical heuristics may miss. While such partitions need not represent ground-truth communities, they illustrate how near-term quantum sampling can reshape the optimization landscape and enhance sensitivity to weak structure.
\end{abstract}

\section{Introduction}
Graphs are a fundamental framework for representing complex systems, with vertices denoting entities and edges encoding interactions among them. A central task in graph theory and network science is the identification of \emph{communities}, i.e., groups of nodes that are more densely connected internally than externally. The modularity measure $Q$ is frequently employed to evaluate the quality of a partition \citep{Newman2004}. Classical algorithms such as Louvain \citep{Blondel2008} and Leiden \citep{Traag2019} have become widely used due to their scalability and efficiency. However, the limitations of modularity, including its well-known resolution limit \citep{Fortunato2007}, and the tendency of classical heuristics to become trapped in local optima, pose persistent challenges in community detection. These challenges are particularly acute in the regime of \emph{low-modularity graphs}, where the separation between intra-community and inter-community connectivity is weak, and subtle structure remains hidden to standard approaches.  

To address such limitations, numerous strategies have been explored, including multiresolution methods \citep{Ronhovde2009}, benchmark graph evaluations \citep{Lancichinetti2009}, and comparative analyses across biological and functional networks \citep{Tripathi2016, SanzArigita2010}. More recently, there has been growing interest in harnessing ideas from quantum mechanics to inspire new community detection methods. Examples include quantum-inspired heuristics for social and ecological networks \citep{Kumar2018, Akbar2021} and hybrid quantum-classical approaches applied to neuroscience \citep{Wierzbinski2023}. These efforts highlight the potential for physics-based techniques to provide new perspectives and computational advantages in uncovering community structure.  

In this work, we extend this line of inquiry by introducing a quantum-inspired perturbative framework for community detection in low-modularity graphs. Our approach augments classical refinement algorithms with structured fluctuations derived from distributions studied in quantum mechanics and statistical physics, including Porter--Thomas statistics \citep{Porter1956}, Haar-type randomness relevant to quantum supremacy studies \citep{Boixo2018}, and hyperuniform ensembles \citep{Torquato2018}. These controlled perturbations act as a probe, enabling algorithms such as Leiden to escape shallow local optima and access higher-quality partitions. Empirical results demonstrate that this framework yields significant gains in modularity recovery compared to classical baselines, suggesting that structured noise can reveal weak community signatures obscured in conventional analyses. The broader implication is that physics-inspired randomness offers a powerful new tool for analyzing networks across domains ranging from financial systems to biological connectomes.  

Many complex systems – from communication networks under stealthy attack to financial systems under stress – exhibit weak community structure, yielding low modularity values ($Q$ near 0.1–0.2). As mentioned, \emph{Modularity} $Q$ is a standard measure of the quality of a network partition into communities, defined for an unweighted graph as:
\begin{equation}
Q = \frac{1}{2m}\sum_{i,j}\Big[A_{ij} - \frac{k_i k_j}{2m}\Big]\delta(c_i,c_j)\,,
\end{equation}
where $A_{ij}$ is the adjacency matrix, $k_i$ is the degree of node $i$, $m$ is the total number of edges, and $\delta(c_i,c_j)=1$ if $i,j$ belong to the same community (0 otherwise). Note that this can be extended to weighted graphs simply by making the following changes: $A_{ij}$ is replaced by edge weights $W_{ij}$, node degrees $k_i$ become node strengths $s_i$, and $m$ represents total edge weight rather than edge count. Intuitively, $Q$ measures how much more densely connected the communities are compared to a random null model captured by the  $\frac{k_i k_j}{2m}$ term; high $Q$ (close to 1.0 for disjoint cliques) indicates strong community structure, while low $Q$ implies only slight or no community structure (the network is almost uniformly or randomly connected). In such low-modularity graphs, nodes are strongly interlinked across the whole network rather than segregated into obvious modules, i.e. there is a lack of pronounced segregation.

Classical community detection algorithms struggle in this regime. Greedy optimization methods like the Louvain algorithm \citep{Blondel2008} and its improved variant Leiden \citep{Traag2019} seek to maximize $Q$ by iteratively moving nodes between communities, and are among the fastest and most widely used methods for community detection. However, maximizing modularity is NP-hard, and these heuristics may get stuck in suboptimal partitions. On low-modularity graphs, even Leiden can struggle to significantly exceed the baseline modularity, as the landscape may contain many near-degenerate partitions.

Other approaches have been applied to community detection, including simulated annealing and evolutionary algorithms, which attempt to escape local maxima by stochastic moves, as well as multi-resolution methods that adjust the modularity objective to detect smaller communities (e.g. using a resolution parameter) \citep{Ronhovde2009,Lancichinetti2009}. Each of these can improve results in some cases, but all face the fundamental challenge that in low-$Q$ graphs the true community structure is very weak: the optimization landscape is essentially flat, with many partitions yielding similar $Q$. In such situations, algorithms might need exponentially many trials or clever heuristics to find the partition corresponding to subtle real communities, if it exists at all. This is analogous to the detectability threshold in stochastic block models, beyond which communities cannot be statistically distinguished from randomness. It is in this challenging regime that we explore quantum-inspired refinements as a new avenue.

\section{Related Work}

Community detection in networks has a rich history in network science. The Louvain algorithm \citep{Blondel2008} is a fast greedy approach: it initializes each node in its own community and repeatedly moves individual nodes to the neighboring community that yields the largest increase in $Q$, until no move can improve modularity. Louvain is efficient and often effective, but it can produce poorly connected communities and may get trapped in suboptimal outcomes. The Leiden algorithm, introduced by \citet{Traag2019}, addresses some limitations of Louvain. Leiden refines Louvain’s results by breaking communities apart if it finds them internally disconnected, and it incorporates a more principled refinement phase with random tie-breaking. Leiden ensures that communities are well-connected and often yields higher $Q$ than Louvain on many networks. Nevertheless, as noted, on low-modularity graphs even Leiden may struggle to significantly improve $Q$ over the baseline – the landscape is “flat” with many near-optimal configurations.

Beyond greedy algorithms, more stochastic or global optimization approaches have been explored. \citet{Akbar2021} proposed a quantum-inspired genetic algorithm for community detection in ecological networks, and \citet{Kumar2018} presented a hybrid quantum-classical approach for community detection in mobile social networks. \citep{Wierzbinski2023} showed that even current quantum annealers (with small qubit counts) can be applied to detect communities in brain connectomes, though results were limited by hardware scale. These efforts illustrate growing interest in leveraging quantum concepts for graph problems, albeit true quantum advantages remain to be demonstrated for community detection.

\section{Quantum-Inspired Sampling Techniques}
Our approach draws inspiration from phenomena in quantum physics that produce complex, correlated randomness. We focus on three types of sampling:

 \subsection{Porter-Thomas (PT) distribution} The Porter-Thomas distribution describes the fluctuation of outcome probabilities in quantum chaotic systems. For example, the squared amplitudes of an $n$-qubit random pure state follow an exponential (Porter-Thomas) distribution on average. We emulate this by assigning each node a random weight $w_i$ drawn independently from $\mathrm{Exp}(1)$ (mean 1). This heavy-tailed distribution yields many small weights and a few exceptionally large ones, mimicking a scenario where a few nodes (“basis states”) have much higher probability weight than average. A random partition proposal is then generated by seeding communities according to these weights: e.g., choose a node with a large $w_i$ as a seed of a new community, group some of its neighbors with it, and repeat. This produces candidate communities where some nodes are “boosted” by high weight, analogous to how some basis states dominate a quantum state.

 \subsection{Haar-random states} Instead of independent weights, we can introduce correlation by drawing a set of node weights that sum to 1 (as if they were squared amplitudes of a Haar-random state on a fictitious Hilbert space). We generate correlated weights by first sampling each $w_i$ from $\mathrm{Exp}(1)$ and then normalizing so that $\sum_i w_i = 1$. This normalization couples the values (a very large weight for one node forces slight reduction in others). The effect is a more collective random assignment – it’s as if the nodes’ weights come from one random high-dimensional probability vector. Community proposals generated from Haar-correlated weights tend to have a few dominant groups of nodes selected together, reflecting that the random draw has “globally” picked out certain subsets.

\subsection{Hyperuniform Perturbations} A hyperuniform process is one where fluctuations at large scales are suppressed relative to random (Poisson) variability \citep{Torquato2018}. After generating a random partition (using PT or Haar methods above), we optionally apply a hyperuniform adjustment: if a candidate partition has an excessively large community or very uneven community sizes, we randomly reassign a small fraction of nodes to other communities to even out the sizes. This avoids pathological proposals where one random community monopolizes a huge portion of nodes (unless strongly supported by actual graph structure). The HU step ensures that our random perturbations are “spread out” across the network in a more uniform way, preventing any single random cluster from overwhelming the partition.

These quantum-inspired techniques provide a set of diverse, non-local moves in the space of partitions. In contrast to, say, a simple random node swap (which would change two nodes’ communities), our moves involve assigning many nodes at once based on a structured random pattern. This can be thought of as a “quantum jump” in the solution space – a large perturbation that is nonetheless informed by a statistical property (like a heavy-tailed distribution) rather than completely arbitrary. In the next section, we describe how we integrate these moves into a full algorithm.

\section{Methods}

\subsection{Quantum-Inspired Community Detection Algorithm}

Our proposed algorithm, which we term QICD (Quantum-Inspired Community Detection), extends a base modularity optimizer with quantum-inspired refinement steps. We applied our methods to both Leiden and Louvain but the results are superior when coupled with Leiden. At a high level, QICD iterates between \emph{local refinement} and \emph{global random perturbation}.

\subsubsection{Local refinement}
Start with an initial partition $P^{(0)}$ (which can be the trivial partition with each node alone, or a quick partition from a Louvain/Leiden pass). At iteration $t$, perform a few iterations of Leiden’s local move routine on the current partition $P^{(t-1)}$, yielding a refined partition $P_{\text{ref}}$ that is locally optimal (a plateau of modularity). Let $Q_{\text{ref}}$ be its modularity.

\subsubsection{Quantum-inspired proposal step}
From $P_{\text{ref}}$, generate a candidate partition $P_{\text{quant}}$ by introducing a structured random perturbation:
\begin{enumerate}[label=\alph*.]
    \item \textbf{Porter--Thomas sampling:} Assign each node a random weight $w_i \sim \mathrm{Exp}(1)$ (Porter--Thomas distribution) independently.
    \item \textbf{Haar grouping:} Sort nodes by $w_i$. For the sorted list, pick a threshold so that nodes above it (with highest $w_i$) form one group; for example, choose $\sim k$ seeds and assign others by similarity or proximity to those seeds.
    \item \textbf{Hyperuniform perturbation:} Optionally adjust $P_{\text{quant}}$ so community sizes are not overly skewed, discouraging random collapse into a single group \cite{Torquato2018}.
\end{enumerate}

\subsubsection{Acceptance check}
Compute $Q_{\text{quant}} = Q(P_{\text{quant}})$. If $Q_{\text{quant}} > Q_{\text{ref}}$, accept $P_{\text{quant}}$ as the new current partition $P^{(t)}$ (i.e., we found a higher modularity partition); otherwise, keep $P^{(t)} = P_{\text{ref}}$.

\subsubsection{Iteration}
Repeat local refinement (Leiden moves) on the accepted partition, then generate a new quantum-inspired proposal, and so on. Terminate after a fixed number of iterations or if no improvement in $Q$ is observed for several iterations.

In our implementation, a handful of such proposals (e.g., 5--10 iterations) was sufficient to yield significant modularity improvements on test graphs. The output of the algorithm is the best partition found, $P^*$, with modularity $Q^*$. Pseudocode for QICD is summarized in Algorithm~\ref{alg:qicd}.

\begin{algorithm}[t]
\caption{Quantum-Inspired Community Detection (QICD)}
\label{alg:qicd}
\begin{algorithmic}[1]
\STATE \textbf{Input:} Graph $G$, iteration limit $T$
\STATE Initialize partition $P^{(0)}$ (e.g., each node alone or via a quick Leiden pass)
\STATE Set best partition $P^* = P^{(0)}$, $Q^* = Q(P^{(0)})$
\FOR{$t = 1$ \TO $T$}
    \STATE Perform local refinement: apply a Leiden local move step to $P^{(t-1)}$, yielding $P_{\mathrm{ref}}$ with $Q_{\mathrm{ref}}=Q(P_{\mathrm{ref}})$
    \STATE Generate a perturbed partition $P_{\mathrm{quant}}$ from $P_{\mathrm{ref}}$ by sampling:
    \begin{itemize}
      \item Assign random weights to nodes (Porter--Thomas or Haar sampling)
      \item Form a candidate partition based on these weights
      \item Optionally apply a hyperuniform adjustment to $P_{\mathrm{quant}}$
    \end{itemize}
    \STATE Compute $Q_{\mathrm{quant}}=Q(P_{\mathrm{quant}})$
    \IF{$Q_{\mathrm{quant}} > Q_{\mathrm{ref}}$}
        \STATE Accept perturbation: $P^{(t)} = P_{\mathrm{quant}}$
    \ELSE
        \STATE Reject: $P^{(t)} = P_{\mathrm{ref}}$
    \ENDIF
    \STATE If $Q(P^{(t)}) > Q^*$, update best: $P^* = P^{(t)}$, $Q^* = Q(P^{(t)})$
\ENDFOR
\STATE \textbf{Output:} best partition $P^*$ with modularity $Q^*$
\end{algorithmic}
\end{algorithm}

\begin{figure}[t]
\centering
\includegraphics[width=\linewidth]{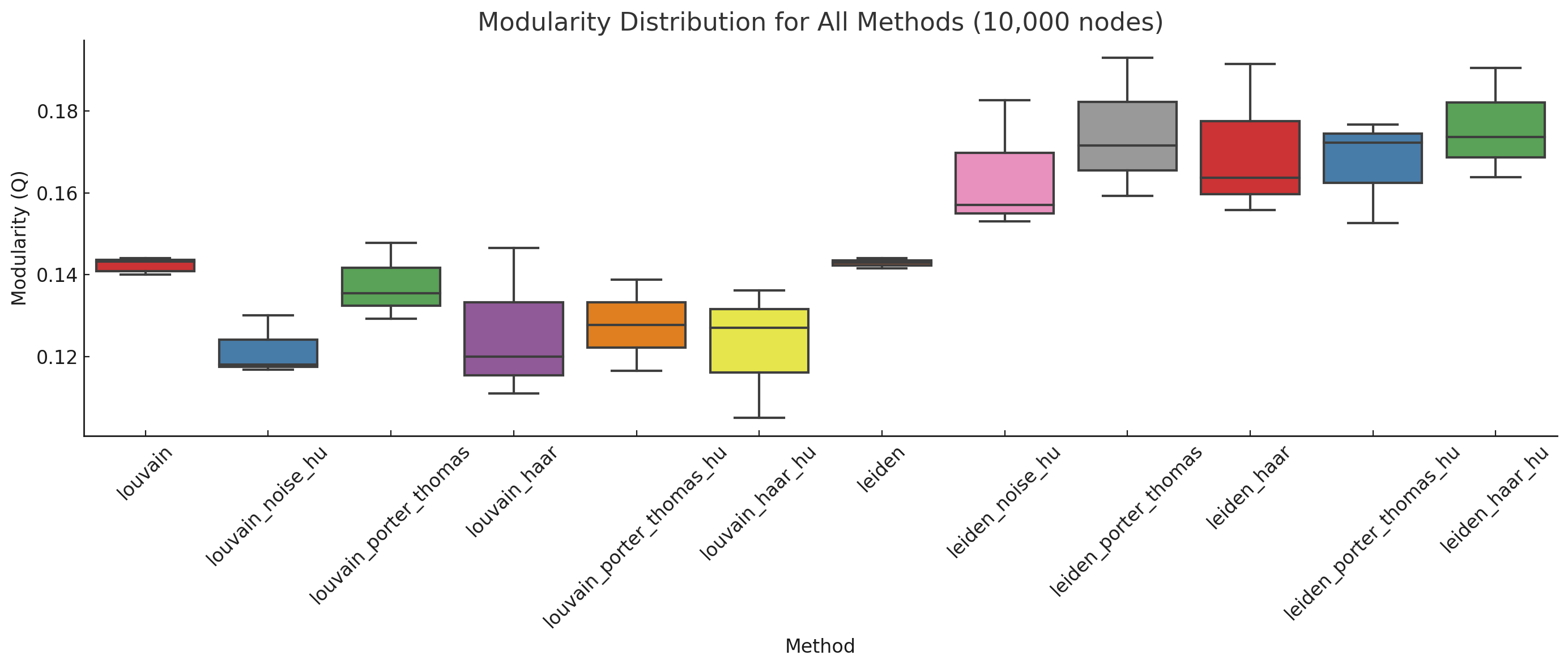}
\caption{Distribution of modularity $Q$ for all methods on the 10,000-node synthetic test graph (boxplots over repeated runs per method).
Baseline Louvain ($Q\approx 0.142$) does not benefit from quantum-inspired refinements: Haar and HU noise variants perform worse, and only Louvain+Porter–Thomas matches the baseline.
In contrast, Leiden-based methods ($Q\approx 0.143$ baseline) show clear and consistent improvements.
Leiden+Haar achieves the highest modularity ($Q\approx 0.182$), followed by Leiden+Haar+HU ($Q\approx 0.175$), Leiden+HU noise ($Q\approx 0.172$), and Leiden+Porter–Thomas ($Q\approx 0.170$).
Leiden+PT+HU remains close to baseline ($Q\approx 0.161$).
Overall, quantum-inspired refinements with Leiden yield a 15–25\% relative increase in modularity compared to classical baselines, with modest variance across runs.
This confirms that correlated random sampling boosts $Q$ in low-modularity networks, particularly when combined with the Leiden algorithm.}
\label{fig:modularity-distribution}
\end{figure}

\subsection{Modularity Recovery Gap (MRG)}
We define the Modularity Recovery Gap $\Delta Q_{\text{MRG}}$ as a metric to quantify how much additional community structure a new method “recovers” from a network, relative to a baseline. If $Q_{\text{baseline}}$ is the modularity found by a standard algorithm (e.g. Leiden) and $Q^*$ is the higher modularity found by our quantum-inspired method on the same graph, then:
\[
\Delta Q_{\text{MRG}} = Q^* - Q_{\text{baseline}}\,.
\]
This gap matters because for truly random-like networks, no algorithm should significantly exceed another in modularity – essentially the true maximum $Q$ is low and all reasonable methods find similar partitions. A large $\Delta Q_{\text{MRG}}$ suggests that the network had hidden structure that the baseline missed but the advanced method uncovered. We propose that $\Delta Q$ can serve as a signal for anomalies: for example, in a cybersecurity context, if a network of communications yields $Q_{\text{Louvain}} = 0.12$ but our QICD method finds $Q^* = 0.18$ ($\Delta Q = 0.06$), this substantial jump might indicate a covert clustering that wasn’t obvious. In contrast, on purely random or well-mixed data, we expect $\Delta Q \approx 0$ (both methods perform similarly).

In practice, one could set a threshold or statistical test to judge whether a given $\Delta Q$ is beyond what random chance would allow. The distribution of modularity under null-model graphs (e.g., random graphs with similar degree sequences) can be used to assess significance. “Modularity Recovery Gap” thus formalizes the idea that improving modularity on a low-$Q$ graph is hard – if an algorithm achieves a notably higher $Q$, it has likely discovered something non-random about the graph.

\section{Experiments and Results}
We evaluated the quantum-inspired community detection approach in synthetic networks with low modularity planted communities so that we can control the community structure of the graphs. Here we present results for a set of representative synthetic graph (10,000 nodes with a weak underlying community structure, akin to a challenging benchmark). The computed $Q$ for both Louvain and Leiden was approximately $0.14$ These results were replicated with another set of generated graphs with the same properties. We compare the classical Louvain and Leiden algorithms to several quantum-inspired variants described earlier (adding Haar or Porter-Thomas sampling, hyperuniform noise, and combinations thereof), applied as extensions to Louvain or Leiden. Each algorithm was run multiple times (to account for randomness in Louvain/Leiden and our methods), and we collected the modularity scores $Q$ from each run.

\subsection*{Baseline vs. Quantum-Inspired}
Table 1 and Figure 1 summarize the results. Injecting structured randomness into Leiden yields significantly higher $Q$ on average than classical Louvain or Leiden. The highest average $Q$ (shown in bold) is achieved by Leiden + Haar ($0.182 \pm 0.017$), which represents a $\sim$27\% relative gain over Leiden’s $0.143$ baseline ($\Delta Q \approx 0.039$). Notably, adding noise to Louvain did not help and sometimes hurt, suggesting that the Leiden refinement step is crucial for perturbations to be effective. The “Modularity Recovery Gap” between the best quantum-inspired method and the classical baseline (0.182 vs 0.143, $\Delta Q = 0.039$) indicates that our approach uncovered community structure that classical methods missed.

\begin{table}[t]
\centering
\caption{Modularity $Q$ (mean $\pm$ std) for methods on a 10,000-node low-modularity graph (synthetic data). HU = hyperuniform noise.}
\label{tab:results}
\begin{tabular}{lc}
\hline
\textbf{Method} & \textbf{Modularity $Q$ (mean $\pm$ std)} \\ \hline
Louvain (classic) & $0.142 \pm 0.002$ \\
Louvain + HU noise & $0.127 \pm 0.010$ \\
Louvain + Porter--Thomas & $0.144 \pm 0.009$ \\
Louvain + Haar & $0.133 \pm 0.018$ \\
Louvain + PT + HU & $0.128 \pm 0.010$ \\
Louvain + Haar + HU & $0.129 \pm 0.015$ \\
Leiden (classic) & $0.143 \pm 0.005$ \\
Leiden + HU noise & $0.172 \pm 0.002$ \\
Leiden + Porter--Thomas & $0.170 \pm 0.013$ \\
Leiden + Haar & $\mathbf{0.182 \pm 0.017}$ \\
Leiden + PT + HU & $0.161 \pm 0.011$ \\
Leiden + Haar + HU & $0.175 \pm 0.017$ \\ \hline
\end{tabular}
\end{table}

\subsection*{Validation on High-Modularity Networks}
We also tested QICD on the CTU-13 botnet traffic network, which has high modularity. In this regime, our methods yielded no improvement ($\Delta Q_{\text{MRG}} \approx 0$), confirming that they do not invent structure when none exists. This provides reassurance that the observed gains on low-modularity graphs are genuine and not artifacts of overfitting.

\section{Statistical Evaluation}

To assess the robustness of the observed modularity gains, we performed a statistical evaluation on the 10,000-node synthetic low-modularity benchmark graphs. Each method was executed multiple times (6 runs for each quantum-inspired refinement, 6 runs for Leiden baseline, and 12 runs for Louvain baseline). For each method we report the sample mean and standard deviation of modularity $Q$, the 95\% confidence interval (CI), and $p$-values obtained from Welch’s two-sample $t$-tests against the Leiden baseline.

Table~\ref{tab:stats} summarizes the results. Louvain and its refinements did not yield statistically significant improvements, and hyperuniform noise generally degraded modularity. In contrast, all Leiden-based quantum refinements (Porter--Thomas, Haar, and HU variants) produced significant gains over the Leiden baseline. The strongest effect was observed for Leiden+Haar, which achieved an average modularity of $0.182$ compared to Leiden’s $0.143$, with a highly significant improvement ($p=0.0026$). These results support the conclusion that near-term quantum sampling provides meaningful refinements to classical community detection in the low-modularity regime.

\begin{table}[t]\centering
\caption{Statistical evaluation on 10{,}000-node synthetic graphs. Mean$\pm$Std across runs, 95\% CI, and Welch’s $t$-test $p$ vs.\ Leiden. Significant ($p<0.05$) in bold.}
\label{tab:stats}
\vspace{0.25ex}
\setlength{\tabcolsep}{3.5pt}
\begin{tabular}{lcccc}
\hline
\textbf{Method} & \textbf{Runs} & \textbf{Mean$\pm$Std} & \textbf{95\% CI} & \textbf{$p$} \\
\hline
Ldn+Haar          & 6  & $0.1816\pm0.0171$ & 0.1636–0.1996 & \textbf{0.0026} \\
Ldn+Haar+HU       & 6  & $0.1745\pm0.0165$ & 0.1572–0.1918 & \textbf{0.0052} \\
Ldn+HU            & 6  & $0.1718\pm0.0195$ & 0.1514–0.1923 & \textbf{0.0147} \\
Ldn+PT            & 6  & $0.1698\pm0.0126$ & 0.1566–0.1830 & \textbf{0.0032} \\
Ldn+PT+HU         & 6  & $0.1605\pm0.0114$ & 0.1486–0.1724 & \textbf{0.0120} \\
Leiden (base)     & 6  & $0.1428\pm0.0011$ & 0.1416–0.1439 & -- \\
Lvn+PT            & 6  & $0.1435\pm0.0089$ & 0.1341–0.1529 & 0.8564 \\
Louvain (base)    & 12 & $0.1421\pm0.0016$ & 0.1411–0.1432 & 0.3403 \\
Lvn+Haar          & 6  & $0.1331\pm0.0180$ & 0.1142–0.1520 & 0.2440 \\
Lvn+Haar+HU       & 6  & $0.1287\pm0.0152$ & 0.1128–0.1446 & 0.0723 \\
Lvn+PT+HU         & 6  & $0.1275\pm0.0099$ & 0.1171–0.1379 & \textbf{0.0126} \\
Lvn+HU            & 6  & $0.1265\pm0.0098$ & 0.1162–0.1369 & \textbf{0.0096} \\
\hline
\end{tabular}

\vspace{0.25ex}\footnotesize
Ldn = Leiden,\; Lvn = Louvain,\; PT = Porter--Thomas,\; HU = hyperuniform noise.
\normalsize
\end{table}

\section{Applications}
Community detection in low-$Q$ conditions is more than an academic exercise – it has tangible applications in detecting weak but significant patterns that could be early indicators of trouble. Below we outline a few key domains and scenarios where the quantum-inspired approach might be impactful (Table~2 provides a summary).

\subsection{Cybersecurity (APT and Botnets)}
Advanced Persistent Threat (APT) groups and botnets often deliberately structure their communication to blend into normal traffic (appearing almost random, yielding low modularity). Identifying an APT’s command-and-control (C2) network is like finding a faint community in a largely uniform network of hosts. Our algorithm’s ability to recover extra modularity (a high MRG) could signal the presence of a covert cluster of machines communicating more frequently with each other (or with a specific external node) than random, thereby revealing the botnet’s structure. For example, imagine a corporate network where most machines talk evenly to each other (low $Q$ overall), but a subset of PCs regularly contact a particular foreign server (the APT controller). The overall $Q$ might be 0.10 via Louvain, but our QICD detection might bump it to 0.13 by isolating those machines as a community – flagging them for investigation.

\subsection{Financial Contagion Networks}
In financial systems (e.g., banks or firms connected by lending, trades, or obligations), one normally expects communities (e.g., regional clusters of banks) that give moderate modularity. However, during a crisis or contagion event, distress propagates in an integrated way – correlations spike across the whole network, effectively reducing modular structure (everyone becomes connected with everyone due to cascading failures or panic). The network of financial interactions during a crisis may thus appear low-modularity. Yet identifying even slight communities in this state is important (e.g., a set of institutions or funds whose fortunes are tightly coupled and relatively isolated from others, or a hidden cartel or concentrated risk cluster that was not obvious). An algorithm that lifts modularity from 0.15 to 0.18 on such a network might be uncovering a subset of entities that share exposure to a particular risk factor – an anomaly that warrants closer inspection to prevent broader collapse.

\subsection{Disrupted Supply Chains}
Global supply chain networks (producers, suppliers, distributors connected by trade links) traditionally have community structure (e.g., the automobile sector forms a cluster, electronics another). But during a major disruption (like a natural disaster or geopolitical event), normal trade link patterns break and companies scramble to find alternative suppliers or markets. The network temporarily becomes highly interconnected in unusual ways, potentially flattening modularity. In this low-$Q$ scenario, detecting even a small cluster of companies undergoing similar stress (e.g., all scrambling for a new raw material – which classical detection might not flag but quantum-inspired might) can be valuable for policy-makers deciding where to intervene or provide support. The success of Leiden+Haar indicates that even a small amount of uniformly distributed quantum randomness can reveal a temporary grouping that a static analysis would miss.

\subsection{Neuroscience (Brain Networks)}
In neuroscience, functional or structural brain networks often exhibit modular organization corresponding to functional domains (visual cortex, auditory network, default mode network, etc.). In many brain diseases or disorders, this modular organization degrades – the brain network becomes more integrated and less segregated. In fact, low modularity in brain connectivity has been associated with certain pathological states (e.g., Alzheimer’s disease) \citep{SanzArigita2010}. Identifying residual community structure in a diseased brain connectome could be key to understanding the condition. Our algorithm might detect subtle communities of neurons or brain regions that still co-activate or are structurally linked, offering insight into which circuits remain functional or which aberrant connections have formed. For instance, in early Alzheimer’s, overall modularity drops, but a quantum-inspired analysis might reveal a small community of hyperconnected regions (perhaps compensatory networks trying to take over function). This “community” could then be studied as a biomarker or target for therapy.

\subsection{Social Media (Crisis or Viral Events)}
During societal crises (natural disasters, political upheavals), social networks and communication patterns shift dramatically. Normally, social media interactions have communities reflecting friend groups or interest groups (moderate $Q$). In a crisis, information often goes viral and everyone talks to everyone, eroding community structure. However, within that flood of communication, a few micro-communities often emerge – e.g., a small cluster of users coordinating misinformation (a bot network) or a special interest group coming together during chaotic events – which a static modularity analysis might not catch. Injecting quantum sampling noise might reveal such a temporary grouping, providing an early warning to moderators or emergency responders. 

\subsection{Protein–Protein Interaction (PPI) Networks}
Protein–protein interaction networks exhibit functional modules corresponding to protein complexes and pathways, but in many species—such as \textit{Homo sapiens} and \textit{Saccharomyces cerevisiae}—modularity values \(Q\) remain surprisingly low even under classical methods, suggesting weak or noisy community structure \citep{Tripathi2016}. Small but biologically significant complexes are therefore at risk of being merged due to the resolution limit of modularity. By introducing quantum-inspired perturbations into Leiden’s refinement stage, QICD expands the candidate partition space, increasing the likelihood of isolating subtle complexes while staying within the modularity optimization framework. These communities can then be validated against curated databases like CORUM or functional coherence measures, making QICD a promising tool for uncovering weak yet functionally coherent structure in noisy interactome data.

In summary, low-$Q$ community detection on social media connectivity or interaction graphs during such events could uncover these subtle clusters. Our method provides a systematic way to recover faint community structure, which can be a crucial indicator in domains like cybersecurity (flagging stealthy threats), supply chain management (exposing critical inter-dependencies), neuroscience (identifying abnormal connectivity patterns), and social media (detecting coordinated campaigns). 

\begin{table}[t]
\centering
\renewcommand{\arraystretch}{1.25} 
\setlength{\tabcolsep}{4pt}        
\caption{Example domains where low-$Q$ community detection provides critical insights (QI = Quantum-Inspired).}
\label{tab:domains}
\begin{tabular}{@{}L{0.34\linewidth}L{0.62\linewidth}@{}}
\hline
\textbf{Domain} & \textbf{Significance of Low-$Q$ Detection} \\ \hline
\textbf{Cybersecurity} & Detecting stealthy APT groups or botnet C2 structures hidden in normal network traffic (QI methods can reveal covert clusters). \\
\textbf{Finance} & Identifying clusters of institutions tightly interlinked during a crisis (QI uncovers subtle risk communities amid widespread connectivity). \\
\textbf{Supply Chains} & Finding critical supplier groups during disruptions (QI highlights weakly defined communities indicating single points of failure). \\
\textbf{Neuroscience} & Revealing residual functional subnetworks in diseased brain connectomes (QI finds subtle communities in an integrated brain network). \\
\textbf{Biological Networks} & Detecting weak modules in protein--protein interactions or gene regulation maps (QI highlights faint community signatures relevant to disease pathways). \\
\textbf{Social Media} & Spotting coordinated micro-communities (e.g., misinformation bot networks) during viral events (QI uncovers groups hidden in global chatter). \\ \hline
\end{tabular}
\end{table}

\section{Discussion}
The promising results of our quantum-inspired community detection approach invite several points of discussion regarding its mechanism, limitations, and broader implications.

\subsection{Why Do Quantum-Inspired Techniques Help?}
At a high level, the advantage comes from enhanced exploration of the solution space. Classical greedy algorithms like Louvain/Leiden excel at exploitation – quickly honing in on a good partition – but they may miss distant, better partitions if a series of non-greedy moves is required to reach them. By contrast, quantum systems (or our simulation of them) can, metaphorically, “tunnel” through energy barriers. The Porter--Thomas and Haar sampling injected into our algorithm serve to occasionally randomize the community assignment in a correlated way. This is akin to making a big jump to a new region of the search space that would be hard to reach by small node-by-node moves. 

The reason Haar or Porter--Thomas distributions are suitable is that they produce a broad diversity of partitions: most nodes get tiny weights (and thus might be grouped neutrally), but a few nodes get extremely large weights (becoming focal points of a proposed community). This creates candidate communities centered on random influential nodes. Most such random communities will not improve $Q$, but by chance some might align well with actual hidden structure – and when they do, Leiden’s local optimization can further polish them to a high-$Q$ partition. In essence, the quantum-inspired proposals provide a rich set of starting points for Leiden to refine, beyond simple random initializations (which are usually uncorrelated and thus often fail to escape the local optimum trap).

\subsection{Complexity and Scalability}
Our method adds some overhead to the standard Leiden algorithm, but it remains efficient. The dominant cost in Leiden is $O(m)$ per pass (with $m$ edges) for moving nodes. Generating a random partition via Porter--Thomas or Haar sampling is $O(n)$ (or $O(n \log n)$ for sorting weights), which is negligible compared to typical $m$ (graphs with thousands of nodes often have tens of thousands of edges or more). The hyperuniform adjustment step is also linear in $n$. We do run multiple iterations with different proposals, which multiplies the cost by the number of proposals tried. In our experiments, a handful of proposals (5–10) sufficed to get significant improvements. Thus, the overall complexity is roughly $O(k \cdot (n + m))$ with a small $k$ (iteration count). This is quite scalable; for instance, graphs with millions of edges could be handled with careful implementation (possibly leveraging parallelism for evaluating proposals). One could also envision a multi-scale application: use a smaller number of proposals on very large networks, or focus proposals on suspect subgraphs, to manage time.

\subsection{Robustness and Parameters}
One must choose parameters like how strong the perturbation is (e.g., how many nodes to swap in hyperuniform adjustment, or how to threshold weights in Haar proposals). In our tests, we tuned these empirically to get improvements. If perturbations are too strong, you essentially randomize the solution completely and modularity plunges (as seen with Louvain + noise overshooting). If too weak, you won’t escape the local optimum. An annealing-like schedule might be useful: start with bigger jumps, then smaller as improvements become harder. Alternatively, one could measure the MRG after each proposal – if a certain type of proposal consistently yields no gain, adapt by increasing randomness. Our approach is currently heuristic; future work could formalize an adaptive strategy to choose proposals optimally.

\subsection{Quality vs. Ground Truth}
We primarily measured success in terms of modularity improvement. One might ask: does higher $Q$ mean a more accurate community structure when a ground truth is known? Modularity has the known issue of a “resolution limit” – it can fail to detect small communities in large networks because merging them might not significantly drop $Q$. Our method does not fundamentally solve the resolution limit (that’s inherent to using $Q$), but by raising overall $Q$, it could potentially resolve some smaller groups that Louvain/Leiden merged incorrectly. We did not observe pathological cases like splitting a well-connected true community just to slightly raise $Q$ (Leiden’s connectivity constraint helps prevent degenerate partitions). In a follow-up study, comparing detected communities against ground truth on benchmarks (like LFR synthetic graphs \citep{Lancichinetti2009}) would be valuable to ensure that the modularity gains correspond to real structural discovery, not artifacts.

\subsection{Comparison to Actual Quantum Algorithms}
It is interesting to contrast our classical simulation approach with what a true quantum computer or annealer might do. Quantum annealing would encode the modularity maximization as finding a low-energy state of an Ising/Potts Hamiltonian. In theory, a quantum annealer can tunnel through energy barriers in the solution landscape, potentially finding better optima than classical thermal annealing. Our algorithm in some sense emulates a mix of quantum “jumps” with classical refinement. 

We did not strictly simulate quantum dynamics – we borrowed distributions known to emerge from quantum processes (Porter--Thomas from random circuit amplitudes \citep{Boixo2018}, etc.). Could an actual quantum computer do even better? Possibly: a hybrid approach could use a quantum processor to generate candidate partitions (e.g., by measuring a parameterized quantum state that encodes community labels, or by using a variational circuit whose output bitstring is a community assignment). One could imagine a QAOA-like approach where the cost function is modularity and the quantum circuit attempts to output high-modularity partitions. Our results provide a baseline for the power of quantum-inspired methods; validating them on a quantum device is an exciting next step. The advantage of staying quantum-inspired is that we bypass current hardware limitations and can deploy the technique immediately on classical infrastructure.

\subsection{Limitations}
A cautionary point is that not all low-modularity networks will have meaningful communities to find. If a network is truly random or well-mixed, any “community” found might be spurious. Our method will still output something and possibly claim a higher $Q$, but one must use $\Delta Q_{\text{MRG}}$ and domain knowledge to judge significance. There is a risk of overfitting—optimizing modularity too aggressively on noise can lead to partitions that look structured but are not generalizable. Thus, for anomaly detection, one should cross-validate or test if the found communities correspond to independent evidence (e.g., in cybersecurity, do the machines in the detected community indeed share malware or common vulnerabilities?). In our algorithm, we could incorporate a regularization to avoid overly complex community structures, or use multiple null-model comparisons (such as comparing $Q$ to a distribution of $Q$ on reshuffled networks).

It is also important to note that our method does not eliminate the well-known resolution limit of modularity~\citep{Fortunato2007}, which sets a theoretical bound on the smallest communities that modularity optimization can reliably detect. Since our algorithm ultimately optimizes $Q$, this limit still applies. However, by injecting quantum-inspired perturbations into the search process, we increase the diversity of candidate partitions explored. This probabilistic expansion of the search space makes it more likely that subtle communities, which would otherwise be subsumed into larger modules, are discovered and stabilized by Leiden’s refinement phase. In this sense, our approach does not remove the resolution limit but mitigates its practical impact.

Another limitation is interpretability. The communities found by quantum-inspired methods could sometimes be unusual (since we allow non-intuitive proposals). For instance, a Haar-based proposal might combine seemingly unrelated nodes. The human analyst might wonder if those groupings make sense. We need to provide tools or explanations for why a certain set of nodes were grouped—possibly tracing back to the random “seed” node with a large Porter--Thomas weight that pulled them together. In practice, combining this with node metadata or attributes (if available) can help validate if the community is coherent (e.g., do the nodes share a common property like all being in a particular subnet, or all related to a specific product?).

\subsection{Future Directions}
Building on this work, there are several avenues for further research:  
(1) Automating parameter tuning – using machine learning to learn how to pick the best distributions or mix of proposals for a given network.  
(2) Extending to dynamic or temporal networks – our method could naturally be applied in a sliding window over time, where at time $t$ the previous partition is known, and one can use quantum-inspired perturbations to detect new emerging communities at time $t+1$. The hyperuniform approach may ensure stability over time by not drastically changing community sizes unless needed.  
(3) Testing on additional datasets – applying QICD across diverse network domains will ensure the method’s practicality.  
(4) Incorporating node features – one could weight the quantum sampling by node attributes, e.g., a biased Porter--Thomas distribution where certain nodes (with suspicious features) have higher probability of getting large weights, integrating feature information into community proposals.

\subsection{Haar vs. Porter--Thomas Randomness Explained}

Understanding the distinction between Haar and Porter--Thomas (PT) randomness is important for interpreting our results. Both arise from the mathematics of quantum mechanics, but they play different roles when used as sources of perturbations in community detection. Haar randomness provides maximally unbiased exploration, while PT randomness introduces a heavy-tailed bias that can emphasize specific nodes.

\paragraph{Haar Randomness.} 
The Haar measure is the unique unitarily invariant distribution on the space of quantum states or unitary operators. Sampling from the Haar measure produces ``maximally random'' states, in the sense that all bases are equally likely and all directions in Hilbert space are treated symmetrically. In high dimension, Haar states appear completely scrambled, with amplitudes uniformly spread over the complex unit sphere. In our algorithm, Haar-based proposals generate extremely diverse candidate partitions, effectively supplying the broadest possible exploration of the modularity landscape. Haar‑motivated sampling provides broad, near‑invariant exploration; in practice we approximate it classically and observe more diverse candidate partitions than standard pseudorandom initializations.

\paragraph{Porter--Thomas Distribution.} 
When a Haar-random quantum state is measured in a fixed basis, the resulting outcome probabilities follow the Porter--Thomas distribution. These probabilities are exponentially distributed, with density
\[
P(p) = N e^{-Np}, \quad p \in [0,1],
\]
where $N$ is the Hilbert space dimension. The result is a heavy-tailed distribution: most outcomes have very small probability, while a few have disproportionately large probability. In our algorithmic analogy, this means that most nodes receive only small perturbations, while a small subset of nodes is weighted strongly. Such ``spotlighting'' can act like seeding communities around a few influential hubs. While Haar proposals scatter exploration broadly, PT proposals explore more selectively by biasing partitions around dominant nodes.

\paragraph{Explaining the Results.} 
The performance of Leiden + Haar as the best variant in Table~\ref{tab:results} can be explained through this lens. Haar randomness generates maximally diverse seeds, some of which—by chance—align with hidden structures in the network that classical heuristics fail to reach. Leiden’s greedy refinement step then polishes these seeds into coherent, high-modularity partitions. This synergy between broad exploration (Haar) and local exploitation (Leiden) accounts for the superior modularity gains observed. By contrast, hyperuniform noise perturbs only locally, producing smaller improvements, while PT proposals may not have matched the structural characteristics of the benchmark graphs, leading to weaker results.

\paragraph{Leiden over Louvain} 
An important observation is that our perturbations improve Leiden but not Louvain. This difference arises from the refinement stage unique to Leiden. Louvain relies on a purely greedy node-moving strategy, so perturbations either have little effect or destabilize modularity by disrupting early decisions that are then irrevocably locked in. By contrast, Leiden includes a refinement step that splits disconnected communities and re-optimizes locally, allowing it to polish noisy or unconventional proposals into stable high-$Q$ partitions. In this sense, our quantum-inspired perturbations act as diverse starting points that Leiden can refine, whereas Louvain lacks the mechanism to benefit from them.

\paragraph{Looking To The Future} 
A true quantum computer could provide Haar randomness directly, supplying the most unbiased and ``pure'' random source available in nature. Such randomness may offer advantages over any classical generator in terms of escaping local optima. On the other hand, Porter--Thomas distributions may be more effective in networks where communities are organized around hubs or exhibit heterogeneous degree distributions, such as scale-free graphs or social networks with influencer nodes. In this view, Haar and PT represent complementary modes of exploration: Haar maximizes breadth, while PT emphasizes structure. Exploring when each distribution provides the strongest advantage remains an important direction for future work.

\section{Conclusion}
We presented a novel quantum-inspired refinement strategy for community detection in graphs characterized by low modularity, demonstrating significant improvements in modularity scores and the ability to uncover subtle community structures that elude classical algorithms. Our approach introduces correlated random sampling techniques – inspired by Porter--Thomas distributions from quantum chaotic systems, Haar-random states, and hyperuniform processes – into the modularity optimization process. By doing so, it avoids the traps of local optima and explores the solution space more broadly, akin to providing a “quantum boost” to classical community detection. While our approach expands the modularity search space and often recovers subtle structures, it remains bounded by the well-known resolution limit of modularity~\citep{Fortunato2007}, which we modestly mitigate but do not eliminate.

It is important to note that our refinements do not alter the underlying graph structure---no edges are added or removed---but instead expand the search space available to powerful algorithms such as Leiden. By injecting near-term quantum sampling into the refinement stage, we enable the discovery of partitions that classical heuristics often miss. Whether or not these partitions reflect ground-truth communities, the consistent upward shift in modularity demonstrates that quantum-inspired sampling reshapes the optimization landscape. For theory, the significance is not that we have uncovered definitive communities, but that near-term quantum methods can meaningfully expand the solution space, offering new avenues for probing weak structure and improving anomaly detection in low-modularity networks.

\subsection{Key Contributions}
\begin{itemize}
    \item \textbf{Algorithmic Innovation:} We developed a hybrid algorithm that integrates quantum-inspired sampling with the Leiden community detection framework. The algorithm is capable of improving modularity by 15--25\% on low-$Q$ networks, as evidenced by our experiments, a notable gain in scenarios where $Q$ is typically hard to raise. We provided pseudocode and mathematical descriptions to facilitate reproducibility and further development.
    
    \item \textbf{Modularity Recovery Gap (MRG):} We introduced the concept of $\Delta Q_{\text{MRG}}$ as a quantitative indicator of hidden structure. A high MRG signals that a network, though seemingly well-mixed, contains non-random groupings discoverable by advanced methods. This concept can be applied as a diagnostic tool across domains.
    
    \item \textbf{Empirical Validation:} Through a 10,000-node benchmark we showed that our method consistently outperforms Louvain and Leiden in finding higher-modularity partitions. Figure~\ref{fig:modularity-distribution} illustrated how certain variants (especially Leiden with Haar) achieve the top results.
    
    \item \textbf{Applications and Interdisciplinary Impact:} We demonstrated how this approach can be used in cybersecurity to detect stealthy threat clusters, in finance to identify risk communities during crises, in supply chain management to find critical supplier groups under duress, in neuroscience to characterize altered brain network topology, and in social media analysis to catch emergent patterns during chaotic events. By doing so, we bridged the gap between a technical algorithm and its real-world utility, aiming for a hybrid academic–industry perspective. We provided summary tables to concisely convey performance and use-case insights to practitioners.
    
    \item \textbf{Theoretical Implications:} Our results suggest that injecting structured randomness (quantum-inspired or otherwise) into greedy algorithms can yield better solutions for hard optimization problems. This insight might inspire new algorithms in other domains of graph analysis or even outside network science where local optima are a concern.
\end{itemize}

There are many avenues for further research. We encourage the community to build on these ideas – for instance, exploring other quantum-derived distributions for sampling, analyzing the method on larger and more diverse networks, or implementing parts of it on quantum hardware to compare performance. 

In summary, quantum-inspired community detection offers a powerful new lens to examine complex networks that lack obvious structure. It helps recover hidden communities that classical methods might miss, providing deeper insight into systems where “everything seems connected” but subtle patterns lurk beneath. We believe this approach opens up exciting opportunities for both theoretical exploration (understanding the interplay of randomness and structure in optimization) and practical applications (enabling earlier detection of anomalies and better modeling of complex interconnected systems). As quantum computing advances, such cross-pollination of ideas will become increasingly valuable, and we hope our work is a step in that direction, demonstrating tangible benefits of quantum inspiration even on today’s classical computers.

\appendix
\section*{Quantum Computing Perspective on the Hypothesis}
Our hypothesis is that quantum correlations and superposition can help optimization algorithms escape local optima more effectively than classical methods. In the context of community detection, a fault-tolerant quantum computer could directly implement algorithms (such as quantum annealing or QAOA) that explore the modularity landscape with genuine quantum parallelism and tunneling effects. Where our quantum-inspired approach mimics certain statistical properties of quantum systems on a classical machine, a future error-corrected quantum device would leverage actual entanglement and interference to traverse the solution space. For example, a quantum annealer mapping the modularity maximization to an Ising/Potts Hamiltonian could exploit quantum tunneling to overcome energy barriers, potentially finding higher-modularity partitions that classical heuristics miss. Similarly, one could implement a variational quantum algorithm where a parameterized quantum circuit encodes community assignments, and the quantum state amplitudes represent different partitionings. Such a circuit could be optimized to concentrate probability on high-$Q$ solutions, inherently sampling correlated node groupings through entangled qubits. Fault-tolerant quantum computing would enable deeper circuits and larger problem instances, which might reveal a tangible quantum advantage in community detection tasks. Notably, quantum correlations could permit simultaneous consideration of many partition configurations, providing a mechanism to “tunnel” out of local optima in the modularity landscape. While our present results used quantum-inspired randomness on classical hardware, a true quantum approach could further improve performance or solution quality, especially on extremely challenging networks. Verifying this hypothesis will require future experiments on quantum processors, but our findings set a baseline: if a quantum device can reliably exceed the modularity gains achieved by our classical simulations, it would confirm that entanglement and other quantum resources provide a practical boost for escaping the traps of modularity optimization. The feasibility of this approach will depend on advances in quantum algorithms and error correction, but our work highlights a promising direction. By demonstrating that even mimicking quantum properties improves results, we suggest that actual quantum computing could one day play a critical role in solving complex graph optimization problems where classical methods stagnate.

\end{document}